# Projecting Hurricane Risk in Atlantic Canada under Climate Change

**Saeed Saviz Naeini[a], Reda Snaiki[a,*], Alejandro Di Luca[b]**

[a] *Department of Construction Engineering, École de Technologie Supérieure, Université du Québec, Montréal, Québec, Canada*

[b] *Département des Sciences de la Terre et de l'Atmosphère, Centre pour l'étude et la Simulation du climat à l'échelle régionale (ESCER), Université du Québec à Montréal, Montréal, Québec, Canada*

[*]*Corresponding author. Email:* reda.snaiki@etsmtl.ca

**Abstract:** Atlantic Canada faces significant hurricane threats from damaging winds and coastal flooding that are projected to intensify under climate change. This study adopts a two-stage framework. First, the evolution of wind and coastal-flood hazards is quantified from a historical baseline (1979–2014) to two future periods: a near future (2024–2059) and a far future (2060–2095). Hazard fields are constructed from large ensembles of physics-informed synthetic hurricane tracks, and changes are evaluated in return-period wind speeds and in inundation depth and extent, with sea-level rise included for flood projections. The second stage estimates hurricane risk using wind as an operational proxy for total loss, combining the simulated wind fields with exposure data and a vulnerability relationship to compute expected damages. This design clarifies how physical drivers change and how those shifts translate into loss potential without requiring fully coupled compound-loss modeling. Results indicate an intensification of wind extremes and a substantial amplification of coastal inundation, yielding higher wind-proxy risk for many coastal communities. Spatial patterns show a heterogeneous escalation of risk concentrated along exposed shorelines and urban corridors. This comprehensive analysis of both hazard evolution and proxy risk provides decision-ready evidence on where and by how much hurricane losses are likely to grow. The approach clarifies the link between physical drivers and loss potential, ensuring compatibility with standard wind-centric workflows used in engineering and insurance practice.

## 1. Introduction

Tropical cyclones, known as hurricanes in the Atlantic basin, rank among the planet's most destructive natural hazards, posing significant threats to coastal populations and infrastructure worldwide [1,2]. Their impacts are multifaceted, driven primarily by intense winds, torrential rainfall, and storm surge which, when combined with astronomical tides and waves, can lead to devastating coastal flooding [3,4]. A growing scientific consensus, documented in reports such as the Sixth Assessment Report of the Intergovernmental Panel on Climate Change [5], confirms that anthropogenic climate change is altering hurricane characteristics [6–9]. While trends in overall frequency remain uncertain, overall projections suggest an increase in the intensity and rainfall rates of the strongest storms, alongside potential poleward shifts in tracks [5,10–15]. These changes are physically linked to rising sea surface temperatures, increased atmospheric moisture content and changes in the general circulation in a warming world [16–18].

Atlantic Canada, with its extensive coastline bordering the North Atlantic, is frequently affected by powerful storm systems, including hurricanes undergoing extratropical transition as they move northward [19–22]. The region's vulnerability to high winds and coastal flooding is amplified by considerable populations, critical infrastructure, and vital economic sectors concentrated in its coastal zones [23–25]. Recent history provides stark reminders of this vulnerability, with devastating impacts from storms like Hurricane Juan (2003), Igor (2010), Dorian (2019), and Fiona (2022), which caused widespread damage across the region [26–29]. In response to this threat, previous studies have begun to assess the potential impacts of climate change on future hurricane hazards [2,6,33–36]. However, many of the prior assessments have often examined one component of the overall threat at a time, such as projecting changes in storm surge, wind speeds, or rainfall rates. Furthermore, while some recent studies have attempted to translate these changing hazards into comprehensive risk assessments, they are often limited by simplified modeling assumptions [17,37,38] or by focusing on other geographical locations [39–42]. Therefore, a dedicated analysis for Atlantic Canada that characterizes the non-stationary nature of both wind and flood hazards, while also providing a quantitative estimate of the total resulting risk, represents a key opportunity for advancing regional understanding.

Given Atlantic Canada's demonstrated vulnerability, understanding how hurricane hazards will evolve under climate change is a critical research priority. Projecting this evolution is complex because it depends on the interplay among changes in storm frequency and intensity across the Atlantic basin, shifts in typical storm tracks, and the dynamics of extratropical transition. Therefore, robust, forward-looking risk assessments are indispensable for clarifying these future threats and enhancing coastal resilience [43–47]. A multi-hazard framing helps capture the compound nature of hurricanes, where high winds and coastal inundation may coincide in space and time; single-hazard analyses can overlook such interactions [48–53]. Consistent, forward-looking assessments therefore benefit from first quantifying changes in both wind and coastal-flood hazards and then evaluating the implications for regional loss potential. Accurately quantifying future risk further requires methodologies that integrate probabilistic hazard modeling under non-stationary climate forcing with appropriate representations of exposure and vulnerability [50,54].

This study provides a quantitative assessment of evolving hurricane risk in Atlantic Canada through a two-stage framework. The analysis first characterizes the non-stationary nature of two primary physical hazards, namely wind and coastal flooding, from a historical baseline (1979-2014) through two future periods representing the near-future (2024-2059) and far-future (2060-2095), using projections from two GCMs. This is achieved by driving wind and coastal flooding hazard models with large ensembles of physics-based synthetic hurricane tracks. The study then quantifies the evolution of total hurricane risk by integrating the projected wind hazard, which serves as a proxy for total impact, with exposure and vulnerability data. The key aims are to quantify the magnitude and spatial distribution of these escalating threats and highlight future risk hotspots. Ultimately, this research provides a forward-looking perspective on Atlantic Canada's evolving risk profile, offering critical evidence for proactive adaptation planning.

## 2. Methodology

### 2.1. Overall framework

The methodology in this study is structured as a two-stage framework: a comprehensive hazard assessment followed by a risk calculation. The first stage characterizes two distinct physical hazards, both driven by the same ensemble of synthetic hurricane events (Sect. 2.2). Wind hazard is modeled through hazard footprints of maximum sustained surface winds (Sect. 2.3.1), while coastal flood hazard is characterized through inundation depth maps generated using a bathtub approach that incorporates projected sea-level rise (Sect. 2.3.2). The second stage quantifies the overall hurricane risk. In this stage, wind intensity serves as an operational proxy for the total hurricane-induced loss. The risk assessment combines the wind hazard footprints with BlackMarble-derived exposure data [55] (Sect. 2.4) and an adapted Emanuel-type vulnerability function [56,57] (Sect. 2.5). This approach allows for a quantitative assessment of total risk alongside a spatial characterization of the two primary evolving hazards. A schematic overview of this workflow is presented in Fig. 1.

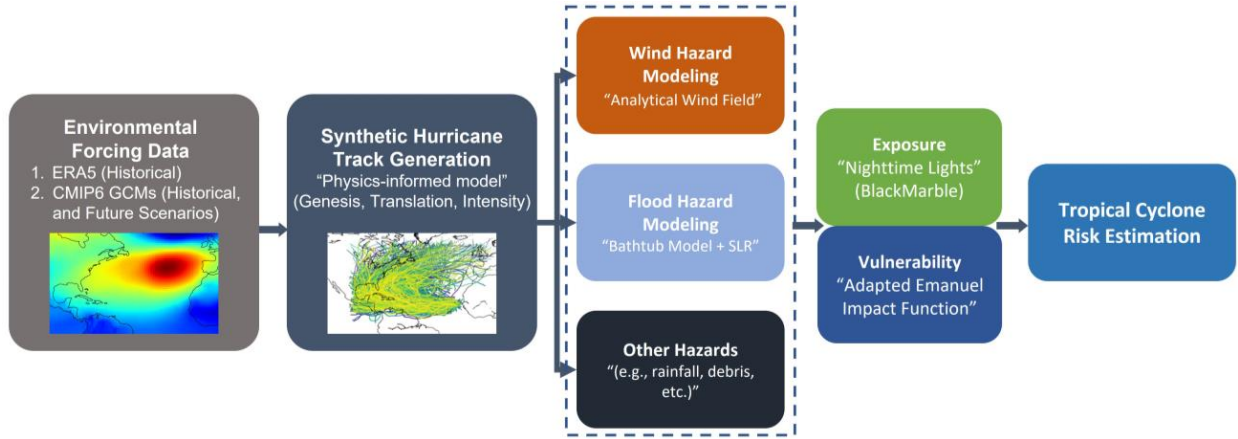

**Fig. 1.** Schematic overview of the proposed risk estimation methodology

## 2.2. Synthetic hurricane track generation

To generate statistically robust hurricane event sets that overcome the limitations of the historical record and allow for analysis under future climate scenarios, a physics-informed synthetic hurricane track model was employed. This approach leverages information from broader geographic regions and climate simulations to populate datasets suitable for estimating the probability of low-frequency, high-impact events. The model simulates the full lifecycle of hurricanes through three core modules: genesis, translation, and intensity. A schematic illustrating this track generation process is presented in Fig. 2.

The genesis module utilizes a stochastic approach, randomly seeding potential storm disturbances across space and time within the simulation domain. These initial seeds are then allowed to evolve based on their interaction with the ambient environmental conditions, mimicking observed patterns of hurricane formation [58]. The translation module governs the storm's trajectory using the beta-and-advection framework [59,60]. The storm's translational velocity ($\mathbf{v}_t$) is calculated at each time step based on the influence of the large-scale environmental winds and a drift component related to the Earth's rotation ($\mathbf{v}_\beta$). The environmental winds ($\mathbf{v}_{850}$ and $\mathbf{v}_{250}$) are the daily-averaged zonal and meridional wind components at the 850-hPa and 250-hPa pressure levels, respectively. In our analysis, wind fields at the storm's location are extracted from either the ERA5 reanalysis or two CMIP6 GCM simulations. The relationship is expressed as:

$$\mathbf{v}_t = (1-\alpha)\mathbf{v}_{250} + \alpha\mathbf{v}_{850} + \mathbf{v}_\beta \cos(\phi) \tag{1}$$

where $\phi$ is the latitude of the storm's center and $\alpha$ is a steering coefficient determining the relative influence of the upper and lower-level winds. The steering coefficient is parameterized as a function of the storm's intensity (maximum wind speed), reflecting the physical principle that stronger storms are steered by a deeper atmospheric layer [60].

For each seeded storm, the intensity module simulates the evolution of the storm's maximum azimuthal wind speed ($v$) based on the FAST (Fast Intensity Simulator) model framework [61]. This simplified physical model uses a coupled system of equations tracking both the maximum wind speed and the inner-core moisture ($m$), incorporating key environmental influences. Specifically, the maximum wind speed ($v$) refers to the peak axisymmetric (rotational) wind at the radius of maximum winds, and the inner-core moisture ($m$) is a non-dimensional bulk variable representing the core's saturation. After asymmetries are added to the model's wind, the final reported value represents a 1-minute sustained wind speed. The core equations are [60]:

$$\frac{dv}{dt} = \frac{1}{2}\frac{C_k}{h}\left[\alpha_o \beta V_p^2 m^3 - (1-\gamma m^3)v^2\right] \tag{2a}$$

$$\frac{dm}{dt} = \frac{1}{2}\frac{C_k}{h}\left[(1-m)v - \chi S m\right] \tag{2b}$$

where $C_k$ is the surface enthalpy exchange coefficient, $h$ is the boundary layer height, $V_p$ is the potential intensity derived from the environmental thermodynamics, $\alpha_o$ relates to ocean interaction, $S$ is the vertical wind shear, and other parameters ($\beta, \gamma, \chi$) depend on environmental thermodynamic properties like entropy deficits and surface humidity [61]. Environmental variables are taken from the background climate fields at the storm location with the following temporal resolutions: daily-mean winds at 850 and 250 hPa (used to compute $S$) and monthly-mean thermodynamic properties, including sea-surface temperature, air temperature, mean sea level pressure, and specific humidity.

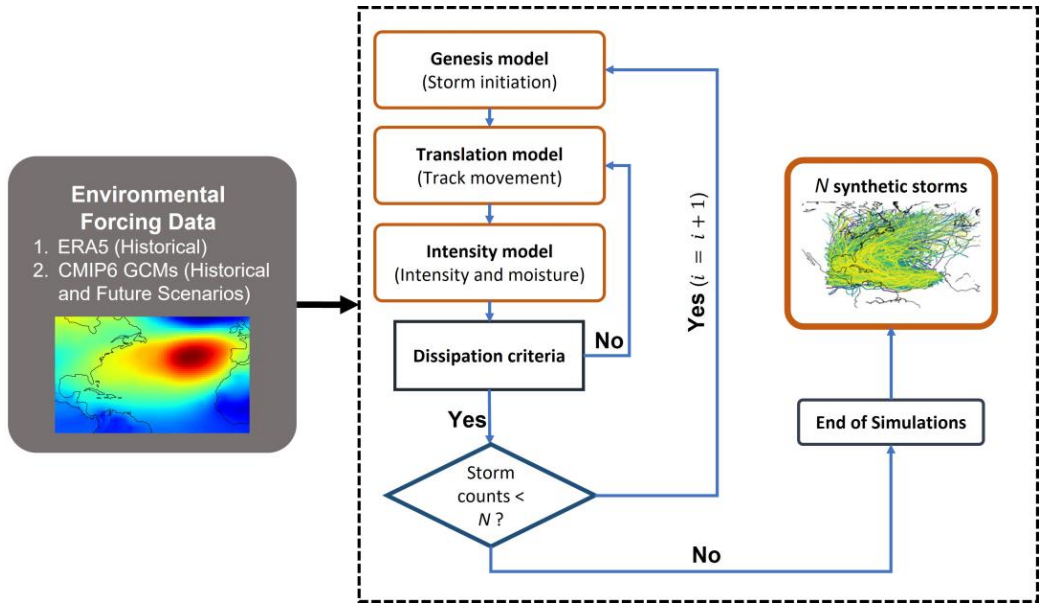

**Fig. 2.** Schematic of the synthetic hurricane track generation process

For the historical reference period (1979-2014), the track and intensity models are forced with data from the ERA5 reanalysis. The model configuration reproduces historical hurricane statistics in line with validations against the IBTrACS database reported by prior studies [60,61]. To generate simulations for future climates, a delta-change approach was employed as a bias correction method to reduce the impact of systematic errors often found in raw GCM outputs. First, climate change perturbations ('deltas') for key environmental fields were calculated using two CMIP6 GCMs (EC-Earth3P-HR and CMCC-CM2-SR5 under the SSP5-8.5 scenario) for the mid-century (2024–2059) and late-century (2060–2095) periods. Thermodynamic fields (e.g., sea surface temperature, air temperature) use monthly-mean deltas (future monthly climatology minus historical monthly climatology), which are added to the monthly ERA5 fields used by the model. Environmental wind fields ($\mathbf{v}_{850}$ and $\mathbf{v}_{250}$) use daily-mean deltas that are added to daily ERA5 winds. The resulting 'perturbed' ERA5 fields serve as the forcing data for the future simulations. The main advantage of this method is that it preserves the realistic and detailed weather patterns of the ERA5 baseline, while using the GCMs only to provide the projected climate change signal. This ensures that the hurricane simulations are not overly influenced by underlying GCM biases [62–67]. This method, often referred to as the pseudo-global warming (PGW) approach, was first introduced by Schär et al. (1996) [66] and has since been widely used to study future changes in tropical and extratropical cyclones [63,67], snowfall over complex terrain [65] and thunderstorms across North America [64]. Using this framework, 10,000 years of synthetic hurricane activity were simulated for each climate period to ensure a statistically robust event set for the subsequent analyses. The output for each synthetic hurricane is an hourly time series detailing its geographical position (latitude, longitude), minimum sea level pressure (MSLP), maximum sustained wind speed ($V_{max}$), and radius of maximum wind ($R_{max}$).

## 2.3. Hazard modeling

Following the generation of synthetic hurricane tracks (Sect. 2.2), the next step is to model the physical hazards associated with each storm event. This process involves two parallel streams: one for wind and one for coastal flooding. First, an analytical wind field model is used to generate a spatial "footprint" of the maximum sustained surface wind speeds for each synthetic hurricane (detailed in Sect. 2.3.1). Second, a bathtub model is employed to simulate the corresponding coastal flood footprint, estimating the maximum inundation depth (detailed in Sect. 2.3.2). This procedure results in a large ensemble of event-based hazard footprints. This ensemble is then statistically analyzed to produce probabilistic hazard maps, which show hazard intensities (e.g., wind speed or flood depth) for various return periods (detailed in Sect. 2.3.3).

### 2.3.1. Wind hazard modeling

The simulation of the hurricane surface wind field for each synthetic event involved several steps. First, the tangential component of the gradient wind ($v_{\theta g}$) was calculated analytically. This component represents the wind speed in the free atmosphere above the boundary layer, balancing pressure gradient, Coriolis, and inertial forces associated with the moving storm. The gradient wind component at a given radius ($r$) and azimuth ($\theta$) is calculated as [68,69]:

$$v_{\theta g} = \frac{1}{2}(-c \cdot sin(\theta - \theta_0) - fr) + \left[\left(\frac{-c \cdot sin(\theta - \theta_0) - f \cdot r}{2}\right)^2 + \frac{r}{\rho} \cdot \frac{\partial P}{\partial r}\right]^{\frac{1}{2}} \tag{3}$$

where $c$ is the storm's translational speed and $\theta_0$ is its direction of motion (obtained from the synthetic track, Sect. 2.2), $f$ is the Coriolis parameter (dependent on latitude $\phi$), $\rho$ is the air density, and $\frac{\partial P}{\partial r}$ is the radial pressure gradient. The pressure gradient term was derived by differentiating the Holland pressure profile model [68], which is defined using storm parameters (MSLP, $R_{max}$, Holland $B$ parameter) from the synthetic track model. The radial component of the gradient wind $(v_{rg})$ was considered negligible, consistent with previous studies [70]. The magnitude of the gradient wind is thus $V_g \approx |v_{\theta g}|$.

Second, the calculated gradient wind speed $(V_g)$ was reduced to estimate the standard 10-meter, 1-minute sustained surface wind speed $(V_s)$ using empirical conversion factors. This step accounts for the frictional effects within the planetary boundary layer, following methodologies used in hurricane risk assessment studies [71]. Specifically, a spatially varying factor based on underlying land cover derived from the World Meteorological Organization (WMO) dataset was used [72]. For this analysis, the land cover is assumed to remain constant across all historical and future time periods. To efficiently generate the wind hazard information, the analysis focused on the impact of storms passing near each location of interest. For each point on a predefined grid, synthetic storm tracks passing within a 250 km radius were identified as potentially influential. For each relevant track segment, the maximum asymmetric surface wind speed at that specific grid point was determined. This location-specific maximum wind speed serves as the primary input for constructing event-based hazard footprints and for the probabilistic hazard assessment described in Sect. 2.3.3.

### 2.3.2. Coastal flood hazard modeling

The coastal flood hazard associated with each synthetic hurricane was assessed using a bathtub modeling approach [73–75], a simplified method that identifies potentially inundated areas by comparing the estimated peak coastal water level during an event against land surface elevation data. The estimation of the peak water level $(\eta_{peak})$ is central to this approach and was determined by summing two key components: the storm surge generated by the hurricane and the projected Sea Level Rise (SLR). Specifically, the storm surge height driven by wind stress and low pressure was estimated for each relevant storm event using an established empirical wind-surge formula [76]. Maximum wind speeds at each coastal location were first computed and then used as inputs to the surge estimation function. To account for future conditions, an SLR value corresponding to each time period (2024-2059 or 2060-2095) was added directly to the estimated storm surge. While relative SLR projections for Atlantic Canada exhibit some spatial variability [8], regionally-averaged values were adopted for this assessment to represent the overall future increase in baseline water levels. Based on high-emission scenario (SSP5-8.5) projections, an SLR of approximately 0.5 meters was used for the near-future period (2024–2059), and 1 meter was used for the far-future period (2060–2095) [8]. The final peak water level was thus calculated as $\eta_{peak} = Storm\ surge + SLR$. The influence of astronomical tides was not explicitly included in this calculation, representing a focus on the storm-induced and climate-change-driven water level components.

The subsequent inundation mapping involved applying the calculated $\eta_{peak}$ for each event to a high-resolution Digital Elevation Model (DEM) representing the coastal topography of the study area. All land grid cells hydraulically connected to the coast and having an elevation lower than the estimated $\eta_{peak}$ are identified as inundated. The inundation depth at each affected grid cell was then computed as $Depth = \eta_{peak} - Ground\ Elevation$. This mapping was performed at a spatial resolution suitable for analysis with the building inventory data.

It is important to acknowledge the inherent assumptions associated with the bathtub model employed in this study. The approach assumes hydrostatic equilibrium (water instantly fills connected areas to the peak level) and neglects hydrodynamic effects such as flow momentum, friction, or the temporal evolution of the inundation. Furthermore, the contributions of wave setup and runup are neglected. The model also treats the landscape as static and does not account for morphological changes during the storm or the presence and potential failure of local flood defense structures unless these features are accurately resolved within the underlying DEM.

The output of this coastal flood hazard modeling stage consists of a set of flood depth footprints, one for each synthetic hurricane event, illustrating the maximum estimated inundation depth across the affected coastal regions within the study area.

### 2.3.3. Probabilistic hazard assessment

From the 10,000 years of simulated hurricane activity, the large ensembles of hazard footprints for both wind (Sect. 2.3.1) and coastal flooding (Sect. 2.3.2) were statistically aggregated to derive probabilistic hazard estimates. This process translates the collection of individual storm impacts into measures of hazard intensity (e.g., wind speed or flood depth) that correspond to specific likelihoods or return periods. For each grid location within the study area and for each hazard type (wind and flood), a time series of the annual maximum intensity experienced over the $n = 10,000$ year simulation period was compiled. These annual maxima were then ranked in descending order, with rank $i = 1$ assigned to the highest intensity, $i = 2$ to the second highest, and so on, up to $i =$ $n$. The annual exceedance probability ($P_e$) for each ranked intensity value was estimated using the empirical Weibull plotting position formula [77,78]:

$$P_e = \frac{i}{n+1} \tag{4}$$

The corresponding return period ($RP$), representing the average time interval between years experiencing an intensity of that magnitude or greater, was calculated as the inverse of the annual exceedance probability:

$$RP = \frac{1}{P_e} = \frac{n+1}{i} \tag{5}$$

Using these relationships, hazard intensity values corresponding to standard return periods of interest (e.g., 50, 100, 300, and 700 years) were determined for each grid location, typically through interpolation between the ranked values. This procedure was applied separately for wind speed and flood depth, and repeated for each climate scenario (historical: 1979-2014; near future: 2024-2059; far future: 2060-2095), allowing for comparison across time periods and scenarios.

The final outputs of the probabilistic hazard assessment stage are spatially explicit maps illustrating the estimated wind speed and flood depth associated with specific return periods (e.g., the 100-year wind speed map and the 100-year flood depth map) for each climate scenario. Hazard curves, plotting intensity against return period or annual exceedance probability for selected locations, can also be generated from this analysis.

## 2.4. Exposure data processing

To quantify the potential impacts from hurricanes, an exposure dataset was processed using BlackMarble data [55]. BlackMarble provides high-resolution estimates of anthropogenic nighttime light intensity, which in this study were used as a spatial proxy for population distribution. The estimated population count in each grid cell was then converted into an economic value by multiplying it by the average provincial gross domestic product (GDP) per capita for the year 2020 [79]. The resulting gridded exposure values, in units of 2020 Canadian dollars (CAD), were formatted to align with the spatial resolution of the wind hazard footprints.

## 2.5. Vulnerability specification

The vulnerability function translates the hazard intensity into a potential damage ratio, forming the critical link between the physical event and its economic impact. For the risk assessment in this study, where wind speed serves as the proxy for total hurricane impact, an adapted vulnerability function based on the work of Emanuel was employed [56,57]. This function relates the maximum sustained surface wind speed at a given location (output from Section 2.3.1) to an expected damage ratio. BlackMarble exposure values (Section 2.4.1) were used to scale the vulnerability outputs, providing a consistent measure of potential impacts across affected assets. The specific functional form used in this study is illustrated in Fig. 3. The Mean Damage Degree (MDD) reflects the severity, while the Percentage of Affected Assets (PAA) indicates the proportion of assets impacted by a given wind intensity. Their product, the Mean Damage Ratio (MDR = MDD × PAA), represents the overall fraction of loss and is the core of the damage calculation [79].

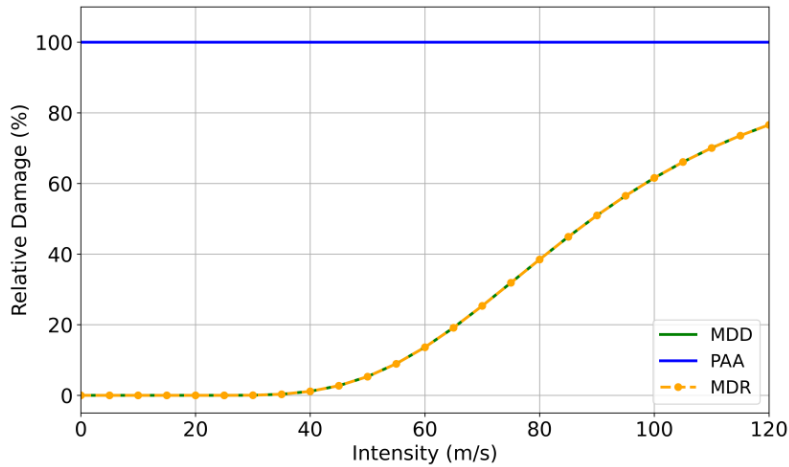

**Fig. 3.** Wind-to-damage vulnerability curves

## 2.6. Risk assessment

The final methodological step is a probabilistic risk assessment to quantify potential economic impacts. The process begins by compiling a full damage history for each individual grid cell. For every storm that impacts a given location over the 10,000-year simulation, the damage is calculated by converting the wind hazard (serving as a proxy for total impact) to a mean damage ratio (MDR) via the vulnerability function and multiplying by the asset value at that location [79]. From this location-specific damage history together with the events' annual occurrence rates, a loss exceedance curve is constructed, which plots the annual probability of exceeding various loss levels. The primary risk metric is the T-year loss, i.e., the loss value on the curve that is exceeded with an annual probability of 1/T. This calculation is performed for every grid cell to produce the spatial risk maps and the return-period plots.

## 3. Case Study

The hazard and risk assessment methodology detailed in Sect. 2 is applied to a case study of Atlantic Canada's coastal regions. This section first outlines the geographic scope and describes the various datasets used for the analysis (Section 3.1). Subsequently, it presents the key findings derived from the simulations (Sect. 3.2), including the characteristics of synthetic hurricane climatology, the projected wind and coastal flood hazards, and the resulting hurricane risk estimates, which are derived using the wind proxy. These results are presented across the different climate scenarios and time periods to illustrate how the hazard probabilities and risk are expected to evolve.

### 3.1. Study area and data

### 3.1.1. Study area description

The study focuses on the coastal regions of Atlantic Canada, a broad area encompassing the provinces of Nova Scotia, New Brunswick, Prince Edward Island, Newfoundland and Labrador, and the Magdalen Islands in Quebec. These regions possess an extensive and complex coastline exposed to weather systems originating in the North Atlantic, including tropical cyclones that often transition into powerful post-tropical storms [80]. Significant portions of the population, critical infrastructure, and economic activities are concentrated in these coastal zones, making them inherently vulnerable to hurricane impacts such as high winds and coastal flooding [47,81,82].

Figure 4 illustrates the geographical extent of the study area where the hazard and risk simulations were conducted. For more detailed analysis, five representative locations were selected from within these regions: Halifax, Nova Scotia (44.60°, -63.47°); St. John's, Newfoundland and Labrador (47.55°, -52.71°); Saint John, New Brunswick (45.26°, -66.04°); Charlottetown, Prince Edward Island (46.22°, -63.14°); and Magdalen Islands, Quebec (47.41°, -61.90°).

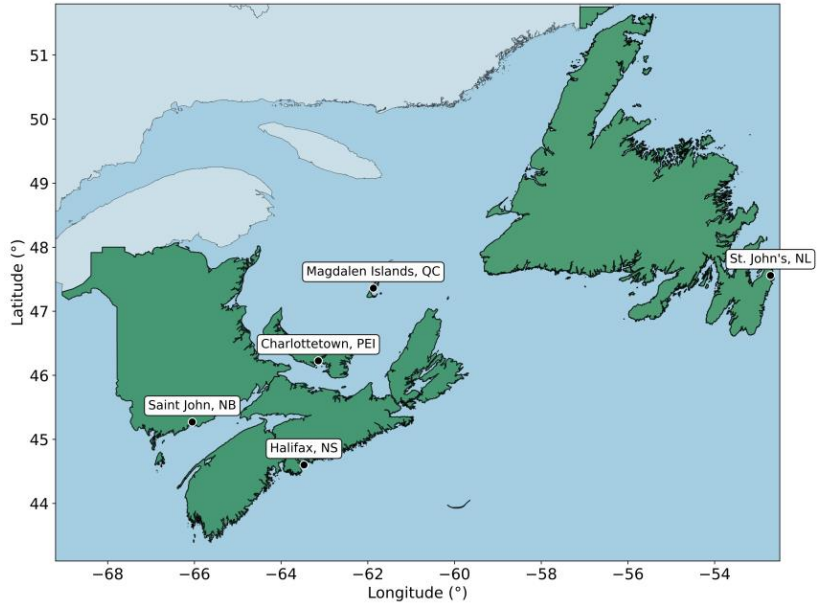

**Fig. 4** Map of the Atlantic Canada study area showing the selected representative locations

### 3.1.2. Data sources

A variety of datasets were employed to drive the synthetic track generation, hazard modeling, exposure representation, and vulnerability assessments detailed in the methodology. The synthetic track model (Sect. 2.2) was driven by environmental conditions from the ERA5 reanalysis database for the historical period (1979-2014) [83]. For future periods (2024-2059 and 2060-2095), corresponding variables were extracted from two GCMs from the CMIP6 project (i.e., EC-Earth3P-HR and CMCC-CM2-SR5) under the SSP5-8.5 high-emission scenario. As mentioned earlier, environmental variables are taken at different temporal resolutions, including daily-averaged zonal and meridional wind components at the 850-hPa and 250-hPa pressure levels, and monthly-mean thermodynamic properties including the sea-surface temperature (sst), air temperature, specific humidity, and mean sea level pressure. Sea level rise projections, incorporated into the flood hazard modeling (Sect. 2.3.2), were adapted from Canadian relative sea-level projections based on the IPCC Sixth Assessment Report under high-emission SSP5-8.5 scenario [8]. Moreover, topographic data essential for the bathtub flood modeling (Sect. 2.3.2) were obtained from the High-Resolution Digital Elevation Model (HRDEM) database from Natural Resources Canada, providing data at a 1 m × 1 m spatial resolution [84].

The economic exposure for the risk assessment was derived from the 2020 NASA BlackMarble dataset, which provides a proxy for population distribution at a 1 km resolution. These population estimates were then converted to an economic exposure value (in 2020 CAD) using provincial GDP per capita data from the World Bank. The vulnerability function, used to translate the wind hazard proxy into a total damage ratio, was adapted from established literature [56]. Other parameters for the hazard models, such as surface wind reduction factors, were also retrieved from published sources [72,76].

### 3.2. Results

### 3.2.1. Synthetic hurricane climatology

The analysis begins with an examination of the synthetic hurricane datasets generated for the historical (1979-2014), near-future (2024–2059), and far-future (2060–2095) periods. The validity of the underlying track generation model is supported by previous validation studies against historical hurricane observations [60,61,84].

To quantify changes in regional hurricane activity, the percentage change in storm frequency passing within a 250-km radius of each location was calculated for future periods relative to the historical baseline. The results, illustrated in Fig. 5, reveal a consistent and notable decrease in storm frequency across all selected locations and under both GCM scenarios. The magnitude of this decrease varies by location and climate model. Projections driven by the EC-Earth3P-HR model show frequency reductions generally ranging from 11% to 18%. The decreases are more pronounced under the CMCC-CM2-SR5 model, with projected declines typically between 18% and 27%. For example, under the CMCC-CM2-SR5 model, the storm frequency at Halifax is projected to decline by 22% in the near future and 25% in the far future. At St. John's, the projected decreases are even larger, at 22% for the near future and 27% for the far future. The specific percentage changes for all locations are detailed in Fig. 5. Overall, these results highlight the non-stationary nature of future hurricane activity, with consistent declines projected across both climate models. The differences in the magnitude of decreases between the two GCMs also emphasize the substantial uncertainty inherent in GCM-driven climate impact assessments.

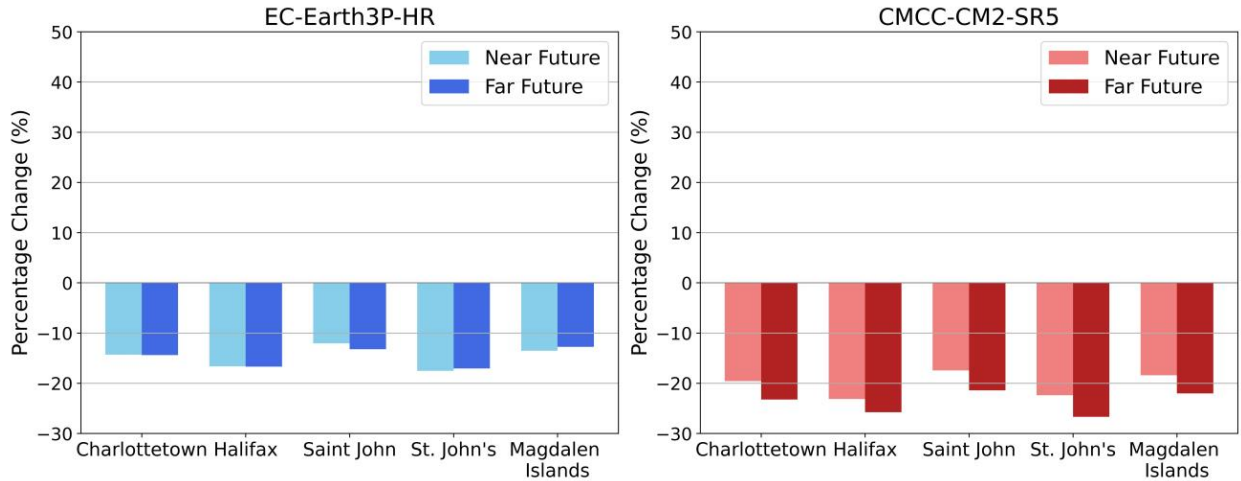

**Fig. 5** Projected percentage change in storm frequency at selected locations for future climate scenarios

### 3.2.2. Hazard analysis

### 3.2.2.1. Wind hazard

Based on the synthetic hurricane climatology, probabilistic hazard levels were computed for across Atlantic Canada. Figure 6 presents maps illustrating the spatial distribution of the 10-meter maximum sustained surface wind speeds for selected return periods (50, 100, 300, and 700 years) using the EC-Earth3P-HR model. A visual analysis reveals a clear and consistent increase in wind hazard for all return periods when progressing from the historical to the near-future and far-future scenarios. This intensification is most pronounced in the far-future period (2060-2095), with the coastal regions of Nova Scotia consistently experiencing the highest wind speeds compared to other parts of Atlantic Canada across all time periods.

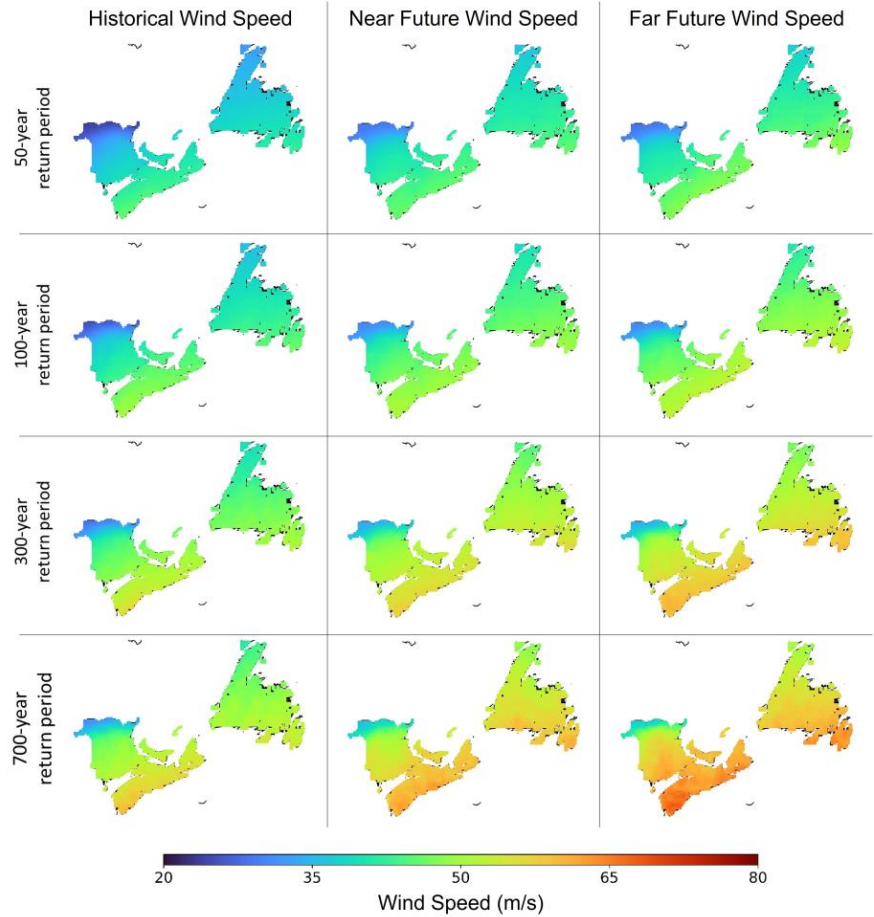

**Fig. 6** Return period wind speeds across Atlantic Canada for historical and future climate scenarios using the EC-Earth3P-HR model

To quantify the magnitude of these changes, the percentage change in the 100-year wind speed was calculated for both future periods relative to the historical baseline, as depicted in Fig. 7. The projections from the EC-Earth3P-HR model are particularly pronounced, indicating that wind speeds for the 100-year event could increase by up to 26% in the far-future scenario, with the most significant changes concentrated along the coast of and Newfoundland, New Brunswick and Nova Scotia. In contrast, the CMCC-CM2-SR5 model produces a more spatially diverse pattern, with both increases and decreases in wind speeds across the region. This pattern includes reductions of up to –7% in parts of Nova Scotia, alongside localized increases reaching as high as 22% in New Brunswick.

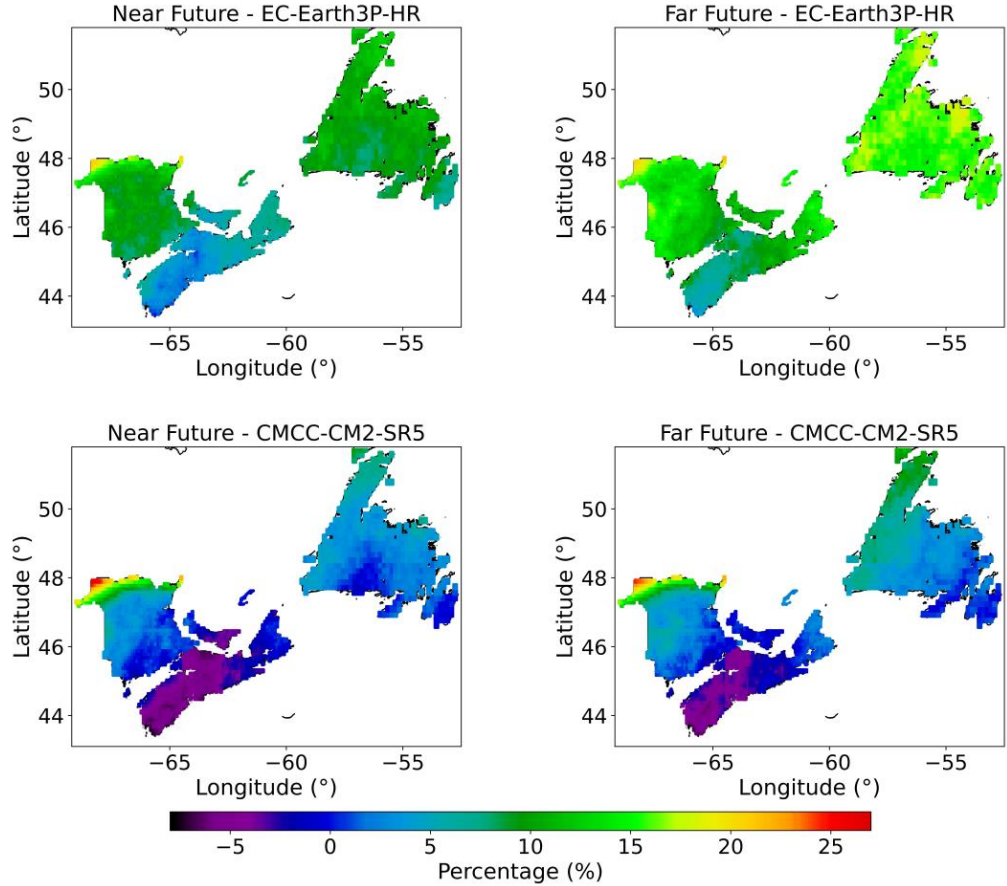

**Fig. 7** Projected percentage change in 100-year wind speed for future scenarios relative to the historical baseline

The trend of intensifying wind hazard is further detailed in the return period plots (Fig. 8) for the five representative locations described in Section 3.1.1. A comparison of the GCM projections at these sites reveals a significant divergence. At the Nova Scotia location, the 100-year wind speed is projected to change by +4% in the near future and +9% in the far future under the EC-Earth3P-HR model. In contrast, the CMCC-CM2-SR5 model projects changes of -6% and -4% for the same periods. In the Magdalen Islands, the 300-year wind speed under EC-Earth3P-HR increases by +9% (near future) and +15% (far future), while the CMCC-CM2-SR5 model shows changes of -2% and +3%. For the Newfoundland site in the far-future, the 300-year wind speed increases by +20% with EC-Earth3P-HR, compared to a +5% increase with CMCC-CM2-SR5. Notably, the EC-Earth3P-HR model consistently projects increases in wind speeds across return periods and locations, whereas the CMCC-CM2-SR5 model shows a more mixed response. This contrast highlights the importance of considering model uncertainty in future risk evaluations.

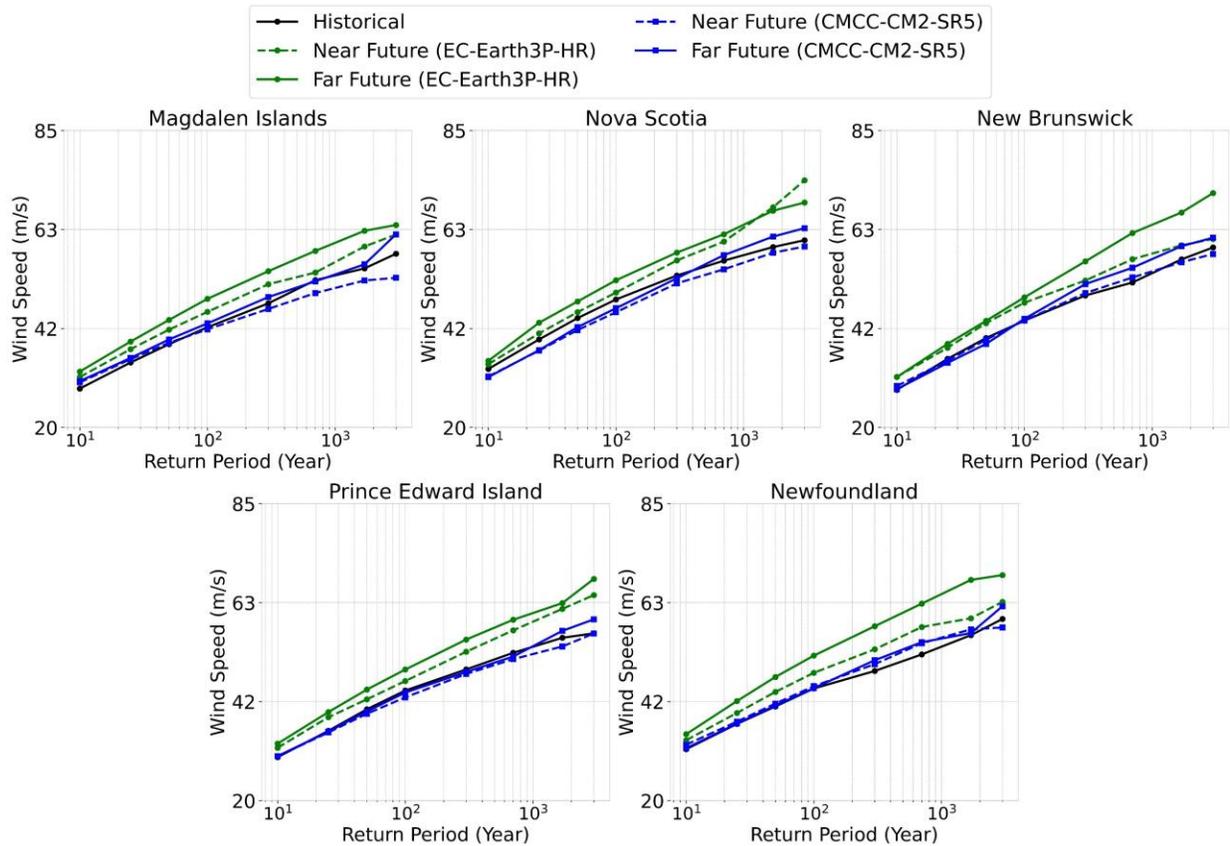

**Fig. 8** Comparison of historical and projected wind speed return periods at selected locations.

### 3.2.2.2. Flood hazard

Probabilistic hazard levels for maximum flood depth were also computed. While the broader study area encompasses all of Atlantic Canada, five representative coastal locations were selected for a detailed, high-resolution analysis to illustrate localized impacts. These sites—located within New Brunswick, Nova Scotia, Prince Edward Island, Newfoundland and Labrador, and Quebec—were modeled at a spatial resolution of approximately 30 by 55 meters. To clearly illustrate the core findings, the flood hazard results presented in this section are driven by the EC-Earth3P-HR model. This focus is used for clarity, as the fundamental trends are dominated by Sea Level Rise and thus are consistent across both GCM scenarios. Figure 9 illustrates the spatial distribution of the 100-year flood depths for three of these areas (Charlottetown, St. John's, and the Magdalen Islands). The maps clearly demonstrate the significant amplifying effect of SLR, which results in a substantial expansion of the inundated area and a notable increase in flood depths compared to scenarios without it.

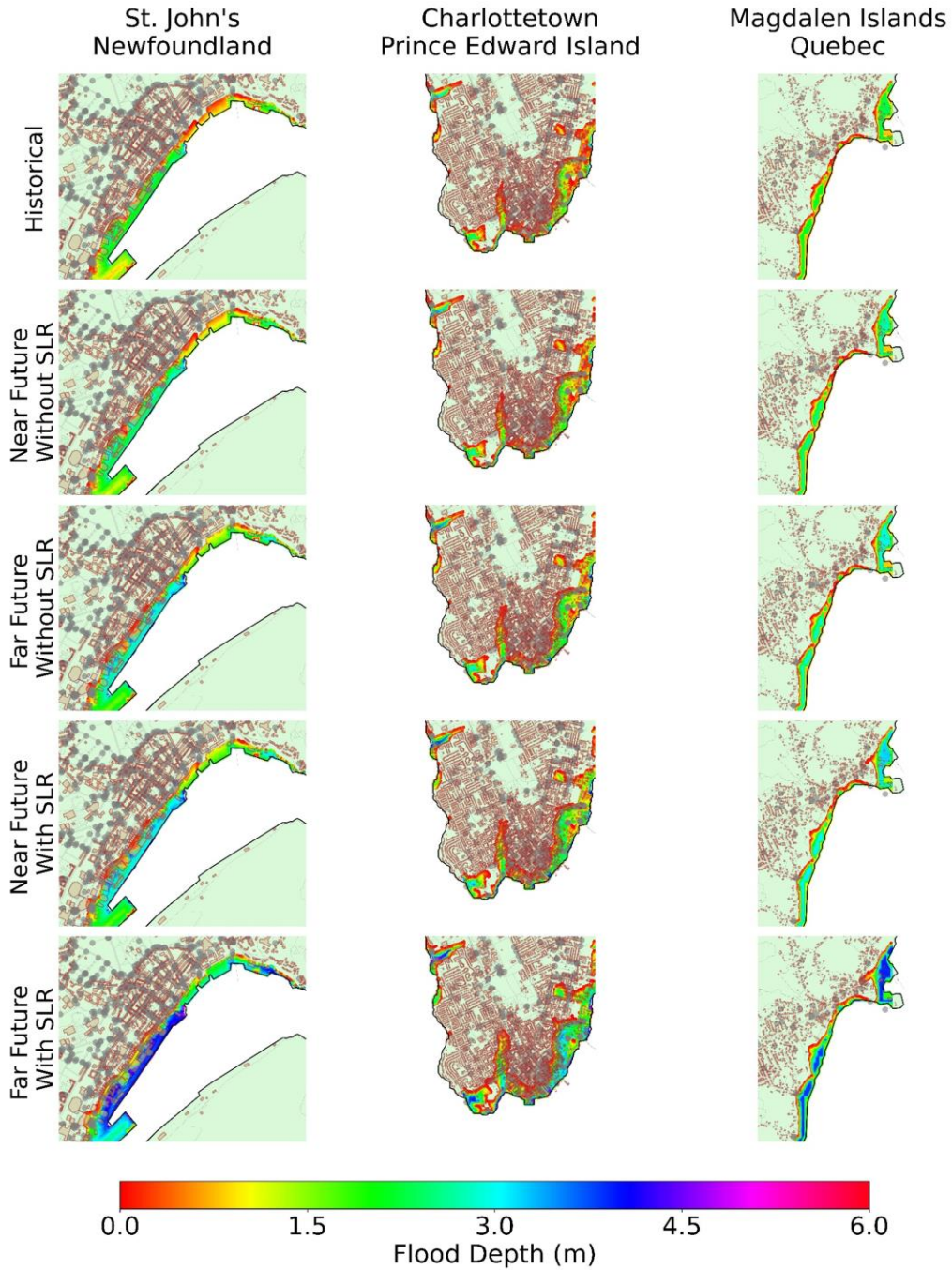

**Fig. 9** Spatial distribution of 100-year coastal flooding for selected locations under historical and EC-Earth3P-HR future scenarios

The changes across all return periods are detailed in Fig. 10, which separates the impact of changing storm characteristics from the combined effect including SLR. The analysis shows that SLR is the dominant factor amplifying future flood hazard. For instance, based on projections from the EC-Earth3P-HR model, at the Halifax site, the 100-year flood depth in the far future is projected to increase by 11% due to storm changes alone; with SLR included, this increase is more than quadrupled to 48%. A similar amplification is seen in the Magdalen Islands, where the 300-year flood depth in the far future increases by 43% from storms alone but jumps to 102% when SLR is added. This pattern of SLR acting as the primary driver is consistent across the other analyzed locations and is also evident in projections from the CMCC-CM2-SR5 model (not shown in Fig. 10 for brevity). For example, under the CMCC-CM2-SR5 model at the Charlottetown site, the 100-year flood depth in the far future is projected to increase by 4% from storm changes alone. With the inclusion of SLR, this increase is quadrupled to 60%. These results confirm that all locations analyzed face a significant and accelerating increase in future coastal flood hazard, driven primarily by SLR.

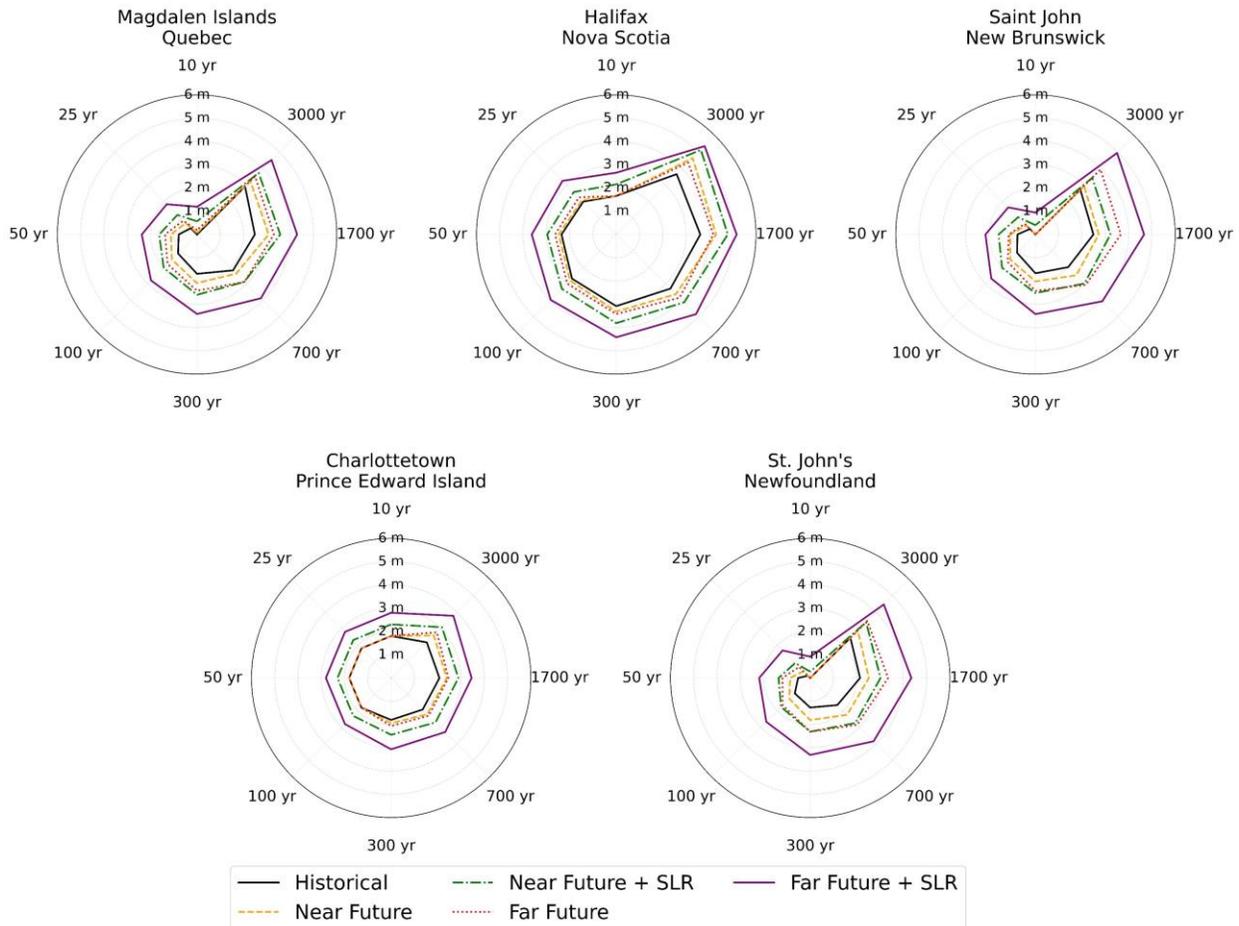

**Fig. 10** Comparison of historical and projected coastal flooding return periods at selected locations, with projections driven by the EC-Earth3P-HR model

### 3.2.3. Risk analysis

The final step of the analysis quantifies the total economic risk from hurricanes. This is achieved by integrating the probabilistic wind hazard, which serves as a proxy for total hurricane impact, with the exposure and vulnerability models. The primary risk metric is Expected Damage ($ED$), which represents the potential loss for events of a given return period. The spatial distribution of $ED$ for a 100-year return period event is illustrated in Fig. 11. The maps show a clear upward trend in risk when comparing the historical scenario to the mid- and late-century periods. The results highlight that the coastal regions of Nova Scotia and Newfoundland are projected to experience the highest absolute damages. This concentration of risk is due to a combination of significant hazard levels and a high concentration of exposed assets in those areas.

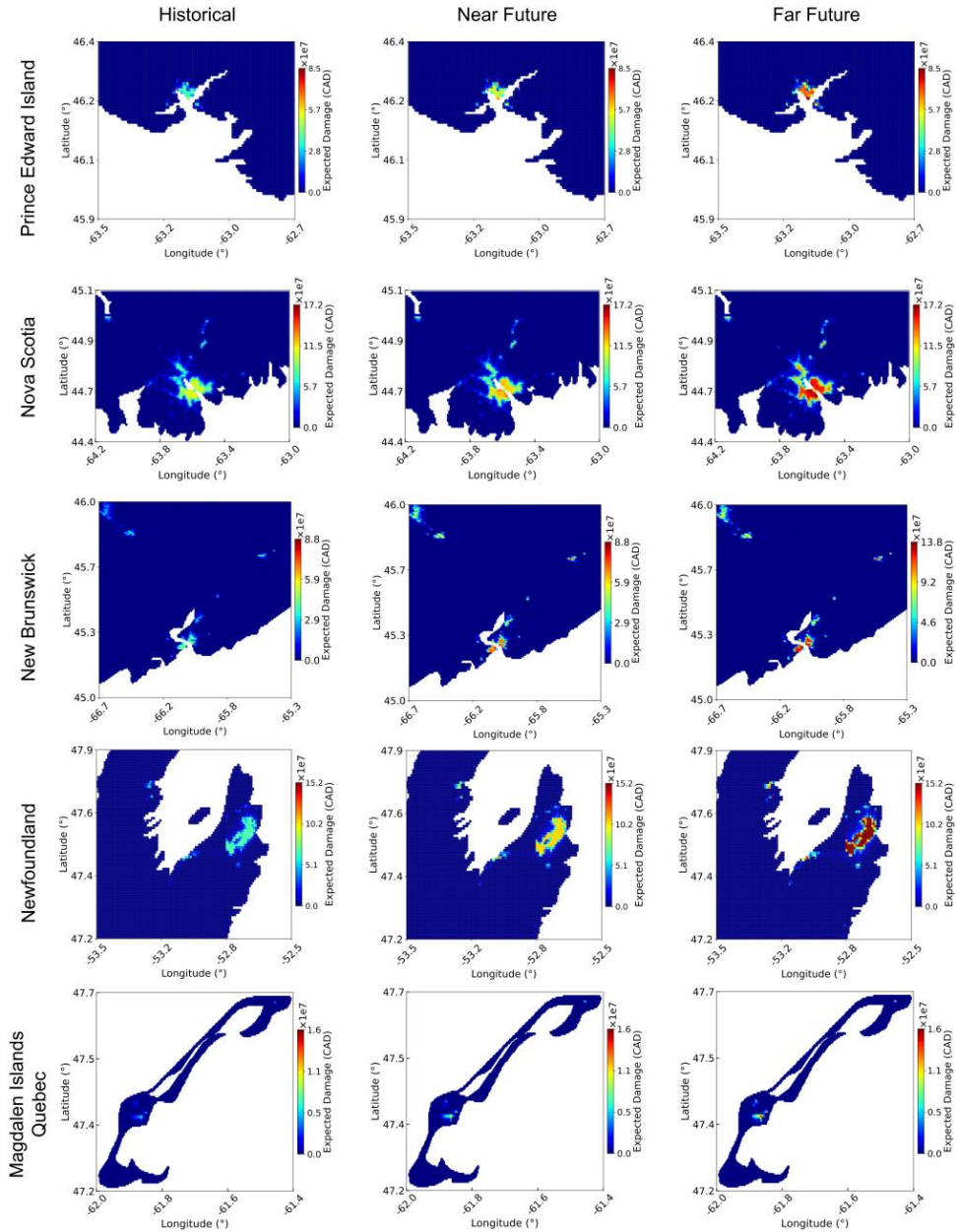

**Fig. 11** Spatial distribution of Expected Damage (*ED*) from the 100-year wind hazard under historical and future scenarios (EC-Earth3P-HR)

Figure 12 provides a more detailed quantification of the increasing risk trends across various return periods for five selected regions. The aggregated damages shown in these plots were calculated by summing the *ED* values for the specific high-exposure sub-areas within each province (as depicted in the Fig. 10 maps). While these aggregated damages generally increase compared to the historical baseline, the plots reveal a notable divergence between the projections from the two GCMs, highlighting significant model uncertainty. For a 300-year event in Prince Edward Island, the far-future scenario shows a 77% increase in *ED* under EC-Earth3P-HR, while the CMCC-CM2-SR5 model projects a 2% decrease. In the Magdalen Islands, damages for the same rare event rise by approximately 102% with EC-Earth3P-HR but by only 23% with CMCC-CM2-SR5. In Nova Scotia, near-future damages rise by 25% under EC-Earth3P-HR but fall by 13% under CMCC-CM2-SR5. These examples show that the EC-Earth3P-HR model consistently produces higher damage projections, whereas the CMCC-CM2-SR5 model produces both increases and decreases depending on the region.

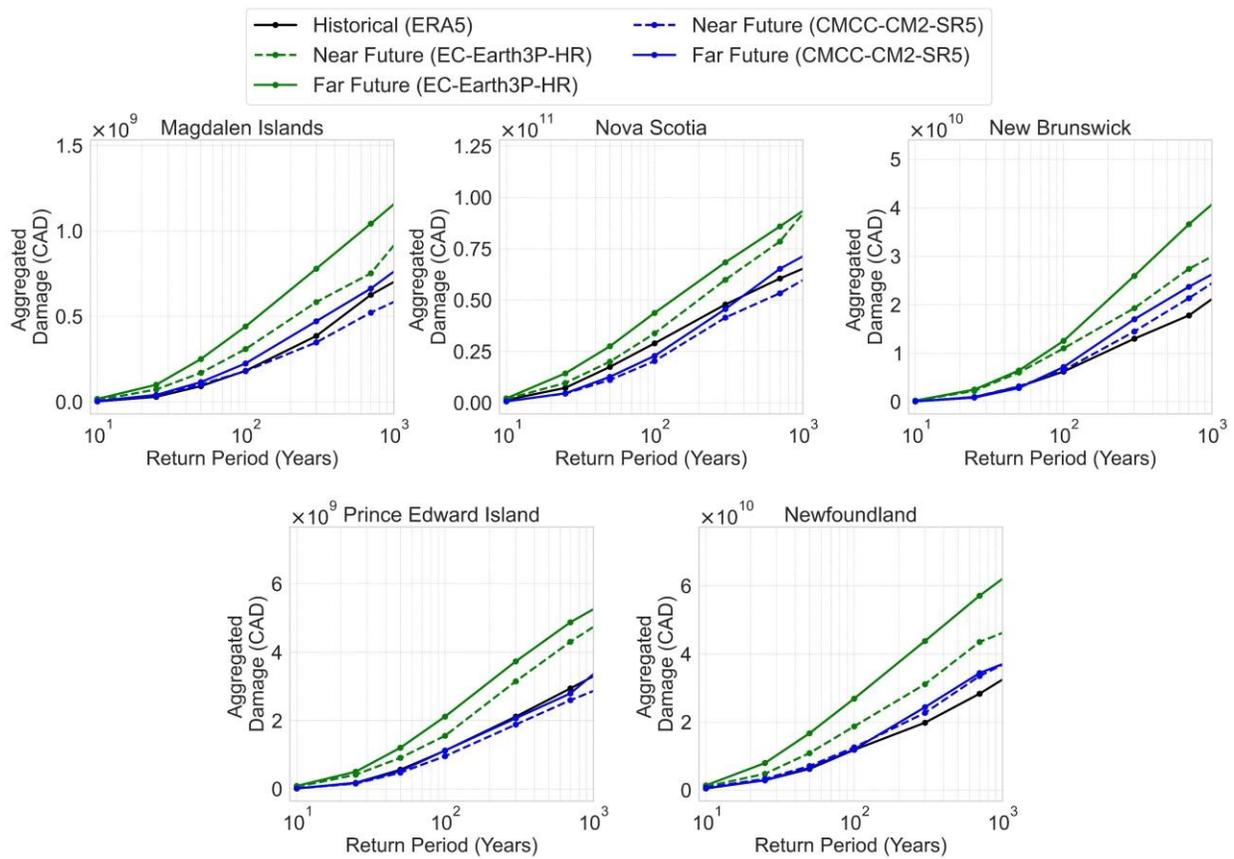

**Fig. 12** Projected Aggregated Damage from wind across various return periods for selected Atlantic Canada regions

## 4. Discussion

This section elaborates on the significance of the results presented in Sect. 3, interpreting the projected changes in hurricane climatology, the associated wind and coastal flood hazards, and the resulting total hurricane risk. It also addresses the implications of the non-stationary nature of these changes, discusses the methodological strengths and limitations, considers the implications for adaptation planning, and suggests directions for future research.

## 4.1. Interpretation of findings

The results of this study indicate a significant projected shift in the hurricane regime affecting Atlantic Canada, leading to a notable escalation in total hurricane risk and an intensification of the coastal flood hazard. A key finding is that future hazard levels will be driven less by the number of storms and more by their intensity. The simulated decrease in the frequency of storms is more than offset by a consistent intensification of wind speeds across all return periods, which in turn amplifies the overall risk calculated using the wind proxy. Spatially, the analysis reveals two distinct hazard profiles: a widespread increase in the wind hazard across the entire coastal region, and a highly concentrated intensification of the flood hazard in low-lying areas, which is dramatically amplified by SLR. Furthermore, the findings reveal two critical dimensions of uncertainty for adaptation planning: the non-stationarity of the climate signal and the divergence between GCMs. The consistent finding that risk and hazard levels are significantly higher in the far-future period than the near-future highlights an accelerating threat, demanding flexible adaptation strategies that can be scaled up over time. The apparent contradiction of decreasing storm frequency alongside increasing hazard underscores the uncertainty in climate projections and reinforces the need for planning approaches that account for a future with fewer but stronger storms, rather than relying on a single prediction.

## 4.2. Methodological considerations and limitations

The methodology employed offers several strengths, including the use of a physics-based synthetic track model to generate large statistical ensembles and the explicit assessment of non-stationarity in future hurricane behavior. However, several limitations and uncertainties must be acknowledged. First, the projections are inherently subject to GCM uncertainty. The results show considerable divergence between the two GCMs, and this study does not encompass the full range of uncertainty from all available CMIP6 models or SSP scenarios. Although a delta-change approach was applied to reduce GCM biases, residual uncertainties from the choice of models and scenarios remain and propagate into the hurricane simulations. Additionally, the choice of bias correction method can also influence the results. This study applied a "delta" to the model inputs as its bias correction strategy. Future work could test the robustness of these findings by comparing them against a statistical approach that applies a "delta" to the model outputs. Second, the hazard models introduce approximations. The wind field calculation relies on parametric relationships, while the bathtub flood model is a significant simplification of flood dynamics, neglecting hydrodynamic effects and wave action. The use of a simplified empirical formula to estimate storm surge from wind speed is a key limitation, as the true relationship is highly sensitive to local bathymetry. The exclusion of astronomical tides is another simplification that could affect peak water levels [85]. Finally, the risk components (exposure and vulnerability) are a major source of uncertainty. A key assumption in this study is that while the hazard evolves with climate change, the socio-economic landscape remains static. Future changes in exposure (e.g., population growth, new coastal development) and vulnerability (e.g., improved building codes or age-related degradation of infrastructure) are not considered. Furthermore, the BlackMarble dataset provides only a coarse proxy for economic exposure, and the vulnerability function, adapted from Emanuel's work, requires significant assumptions about its applicability to the specific building archetypes in Atlantic Canada. In this context, while the wind-proxy approach provides a useful baseline for regional risk evaluation, it is not sufficient for capturing the full spectrum of potential coastal damages, particularly from flooding. A more robust assessment would require the integration of flood-specific fragility curves and high-resolution hazard data, applied at the building level. This highlights that while wind serves as a valuable proxy for large-scale risk quantification, a detailed, flood-specific risk assessment is needed to capture the most severe localized impacts. Lastly, potential interactions between hazards (e.g., wind damage increasing a building's vulnerability to flooding) and other processes, such as rainfall-induced compound flooding, were not considered in this analysis.

### 4.3. Implications for adaptation and research priorities

Projected increases in wind-driven loss potential and coastal-flood hazard, together with clear non-stationarity, call for adaptation that looks beyond historical baselines toward forward-looking planning. Land-use policy and infrastructure design should incorporate elevated future wind loads and flood levels, including sea-level rise, across relevant design lifetimes. Risk hotspot maps can guide zoning by limiting new development in the most exposed areas and by elevating construction standards in zones with increasing risk. Disaster risk reduction should be multi-hazard, combining improved early warning, evacuation and sheltering strategies that reflect the different spatial footprints of wind and flooding. Community programs, emergency planning, and asset management should be informed by the projected geographic shifts in exposure and loss potential. Integrated and adaptive pathways are essential. Planning should address wind and flood jointly, recognizing that the dominant driver varies by location and may evolve over time. Given uncertainty and potential acceleration, adaptation plans should be flexible and iterative, with predefined monitoring indicators and trigger points for updating standards, investments, and land-use decisions as new information becomes available.

Several research priorities follow from these findings. First, hazard characterization would benefit from higher-resolution climate inputs, improved downscaling and bias-correction methods, and for coastal flooding, replacement of the bathtub approach with hydrodynamic modeling that resolves timing, velocities, and wave effects. Second, exposure and vulnerability components require refinement through region-specific vulnerability functions for Atlantic Canada building archetypes, calibrated with local claims and inspection data, and through a consistent exposure database usable for both wind and flood analyses. Third, the scope of risk assessment should expand to compound events, such as storm surge coincident with heavy rainfall, and to dynamic interactions where wind damage alters subsequent flood vulnerability. Embedding specific adaptation measures, including seawalls, building elevation, critical-asset hardening, and zoning changes, directly within the risk-modeling workflow would enable comparative effectiveness analysis. Fourth, future work should incorporate changes in exposure and vulnerability over time, reflecting urban growth, asset value escalation, demographic shifts, infrastructure aging, retrofit and code adoption rates, and managed retreat. This can be implemented through inventory-evolution modules linked to socioeconomic pathways and time-stamped fragility functions that update with materials, design standards, and adaptation uptake. Finally, broader multi-model ensembles across GCMs and SSPs, coupled with explicit uncertainty quantification, can provide more robust decision support for design codes, insurance pricing, and public investment.

## 5. Conclusion

This study presented a quantitative assessment of evolving hurricane threats in Atlantic Canada, characterizing future changes in both wind and coastal flood hazards and estimating the resulting total hurricane risk using a wind-proxy methodology. The findings reveal a complex and accelerating threat profile. Projections point to a future with fewer but more intense hurricanes, leading to a widespread increase in wind hazard, particularly along the coasts of Nova Scotia and Newfoundland. For coastal flooding, the analysis demonstrates that Sea Level Rise is the dominant driver of escalating hazard, dramatically amplifying the effects of stronger storms. The study highlights significant uncertainty in the magnitude of future risk due to divergent outcomes from the climate models. However, the spatial analysis consistently finds that the highest absolute damages are concentrated in coastal communities with significant exposed assets. While a full, separate risk assessment for flooding was beyond the scope of this work, the results clearly illustrate the dual threats facing the region. Relying on the wind-proxy approach, a standard practice in large-scale risk assessment, provides a crucial estimate of escalating economic damages. Ultimately, this study provides a quantitative, forward-looking assessment of Atlantic Canada's evolving hurricane risk profile. By demonstrating a shift toward fewer but more intense storms, highlighting the dominant role of Sea Level Rise in amplifying coastal flooding, and identifying future risk hotspots, the findings provide critical, actionable evidence for developing proactive and spatially explicit adaptation strategies to build resilience against a non-stationary, multi-hazard threat.

## Declaration of Competing Interest

The authors declare that they have no known competing financial interests or personal relationships that could have appeared to influence the work reported in this paper.

## Acknowledgements

This work was supported by the Natural Sciences and Engineering Research Council of Canada (NSERC) [grant number CRSNG RGPIN 2022-03492] and AdapT – Institut de recherche sur les infrastructures résilientes et circulaires.

## References

1. Almeida, L. S., Goerlandt*, F., & Pelot, R. (2019). Effects of major hurricanes in Atlantic Canada from 2003 to 2018. In *Risk Analysis Based on Data and Crisis Response Beyond Knowledge*. CRC Press.
2. Mendez-Tejeda, R., & Hernández-Ayala, J. (2023). Links between climate change and hurricanes in the North Atlantic. *PLOS Climate*, 2, e0000186. https://doi.org/10.1371/journal.pclm.0000186
3. Naeini, S. S., & Snaiki, R. (2024). A novel hybrid machine learning model for rapid assessment of wave and storm surge responses over an extended coastal region. *Coastal Engineering*, *190*, 104503. https://doi.org/10.1016/j.coastaleng.2024.104503
4. Nofal, O. M., Amini, K., Padgett, J. E., van de Lindt, J. W., Rosenheim, N., Darestani, Y. M., Enderami, A., Sutley, E. J., Hamideh, S., & Duenas-Osorio, L. (2023). Multi-hazard socio-physical resilience assessment of hurricane-induced hazards on coastal communities. *Resilient Cities and Structures*, *2*(2), 67–81. https://doi.org/10.1016/j.rcns.2023.07.003

5.  Seneviratne, S. I., Zhang, X., Adnan, M., Badi, W., Dereczynski, C., Di Luca, A., Ghosh, S., Iskandar, I., Kossin, J., Lewis, S., Otto, F., Pinto, I., Satoh, M., Vicente-Serrano, S. M., Wehner, M., & Zhou, B. (2021). Weather and Climate Extreme Events in a Changing Climate. In MassonDelmotte, V., P. Zhai, A. Pirani, S.L. Connors, C. Péan, S. Berger, N. Caud, Y. Chen, L. Goldfarb, M.I. Gomis, M. Huang, K. Leitzell, E. Lonnoy, J.B.R. Matthews, T.K. Maycock, T. Waterfield, O. Yelekçi, R. Yu & B. Zh (Eds.), *Climate Change 2021: The Physical Science Basis. Contribution of Working Group I to the Sixth Assessment Report of the Intergovernmental Panel on Climate Change* (pp. 1513–1766). https://doi.org/10.1017/9781009157896.013

6.  Balaguru, K., Xu, W., Chang, C.-C., Leung, L. R., Judi, D. R., Hagos, S. M., Wehner, M. F., Kossin, J. P., & Ting, M. (2023). Increased U.S. coastal hurricane risk under climate change. *Science Advances*, *9*(14), eadf0259. https://doi.org/10.1126/sciadv.adf0259

7.  Kossin, J. P. (2018). A global slowdown of tropical-cyclone translation speed. *Nature*, *558*(7708), 104–107. https://doi.org/10.1038/s41586-018-0158-3

8.  Lee, H., Calvin, K., Dasgupta, D., Krinner, G., Mukherji, A., Thorne, P., Trisos, C., Romero, J., Aldunce, P., Barret, K., Blanco, G., Cheung, W. W. L., Connors, S. L., Denton, F., Diongue-Niang, A., Dodman, D., Garschagen, M., Geden, O., Hayward, B., … Park, Y. (2023). *IPCC, 2023: Climate Change 2023: Synthesis Report, Summary for Policymakers. Contribution of Working Groups I, II and III to the Sixth Assessment Report of the Intergovernmental Panel on Climate Change [Core Writing Team, H. Lee and J. Romero (eds.)]. IPCC, Geneva, Switzerland.* [Monograph]. Intergovernmental Panel on Climate Change (IPCC). https://doi.org/10.59327/IPCC/AR6-9789291691647.001

9.  Walsh, K. J. E., McBride, J. L., Klotzbach, P. J., Balachandran, S., Camargo, S. J., Holland, G., Knutson, T. R., Kossin, J. P., Lee, T., Sobel, A., & Sugi, M. (2016). Tropical cyclones and climate change. *WIREs Climate Change*, *7*(1), 65–89. https://doi.org/10.1002/wcc.371

10. Gori, A., Lin, N., Xi, D., & Emanuel, K. (2022). Tropical cyclone climatology change greatly exacerbates US extreme rainfall–surge hazard. *Nature Climate Change*, *12*(2), 171–178. https://doi.org/10.1038/s41558-021-01272-7

11. Lin, I.-I., Camargo, S. J., Lien, C.-C., Shi, C.-A., & Kossin, J. P. (2023). Poleward migration as global warming's possible self-regulator to restrain future western North Pacific Tropical Cyclone's intensification. *Npj Climate and Atmospheric Science*, *6*(1), 34. https://doi.org/10.1038/s41612-023-00329-y

12. Reed, K. A., Wehner, M. F., & Zarzycki, C. M. (2022). Attribution of 2020 hurricane season extreme rainfall to human-induced climate change. *Nature Communications*, *13*(1), 1905. https://doi.org/10.1038/s41467-022-29379-1

13. Snaiki, R., & Parida, S. S. (2023). A data-driven physics-informed stochastic framework for hurricane-induced risk estimation of transmission tower-line systems under a changing climate. *Engineering Structures*, *280*, 115673. https://doi.org/10.1016/j.engstruct.2023.115673

14. Studholme, J., Fedorov, A. V., Gulev, S. K., Emanuel, K., & Hodges, K. (2022). Poleward expansion of tropical cyclone latitudes in warming climates. *Nature Geoscience*, *15*(1), 14–28. https://doi.org/10.1038/s41561-021-00859-1

15. Tu, S., Chan, J. C. L., Xu, J., Zhong, Q., Zhou, W., & Zhang, Y. (2022). Increase in tropical cyclone rain rate with translation speed. *Nature Communications*, *13*(1), 7325. https://doi.org/10.1038/s41467-022-35113-8

16. Emanuel, K. (2005). Increasing destructiveness of tropical cyclones over the past 30 years. *Nature*, *436*(7051), 686–688. https://doi.org/10.1038/nature03906

17. Knutson, T. R., Sirutis, J. J., Zhao, M., Tuleya, R. E., Bender, M., Vecchi, G. A., Villarini, G., & Chavas, D. (2015). *Global Projections of Intense Tropical Cyclone Activity for the Late Twenty-First Century from Dynamical Downscaling of CMIP5/RCP4.5 Scenarios*. https://doi.org/10.1175/JCLI-D-15-0129.1

18. Salarieh, B., Ugwu, I. A., & Salman, A. M. (2023). Impact of changes in sea surface temperature due to climate change on hurricane wind and storm surge hazards across US Atlantic and Gulf coast regions. *SN Applied Sciences*, *5*(8), 205. https://doi.org/10.1007/s42452-023-05423-7

19. Garin, A., Pausata, F. S. R., Boudreault, M., & Ingrosso, R. (2024). The impacts of climate change on tropical-to-extratropical transitions in the North-Atlantic basin  *EGUsphere*, 1–24. https://doi.org/10.5194/egusphere-2024-3435

20. Jung, C., & Lackmann, G. M. (2023). Changes in Tropical Cyclones Undergoing Extratropical Transition in a Warming Climate: Quasi-Idealized Numerical Experiments of North Atlantic Landfalling Events. *Geophysical Research Letters*, *50*(8), e2022GL101963. https://doi.org/10.1029/2022GL101963

21. Plante, M., Son, S.-W., Atallah, E., Gyakum, J., & Grise, K. (2015). Extratropical cyclone climatology across eastern Canada. *International Journal of Climatology*, *35*(10), 2759–2776. https://doi.org/10.1002/joc.4170

22. Zadra, A., McTaggart-Cowan, R., Vaillancourt, P. A., Roch, M., Bélair, S., & Leduc, A.-M. (2014). *Evaluation of Tropical Cyclones in the Canadian Global Modeling System: Sensitivity to Moist Process Parameterization*. https://doi.org/10.1175/MWR-D-13-00124.1

23. Li, S. H. (2023). Effect of nonstationary extreme wind speeds and ground snow loads on the structural reliability in a future Canadian changing climate. *Structural Safety*, *101*, 102296. https://doi.org/10.1016/j.strusafe.2022.102296

24. Pang, T., Shah, M. A. R., Dau, Q. V., & Wang, X. (2024). Assessing the social risks of flooding for coastal societies: A case study for Prince Edward Island, Canada. *Environmental Research Communications*, *6*(7), 075027. https://doi.org/10.1088/2515-7620/ad61c8

25. Vasseur, L., Thornbush, M. J., & Plante, S. (2022). Engaging Communities in Adaptation to Climate Change by Understanding the Dimensions of Social Capital in Atlantic Canada. *Sustainability*, *14*(9), Article 9. https://doi.org/10.3390/su14095250

26. Aziz, E. (2021). Key lessons from Hurricane Dorian: The benefits of a flexible top-down storm response. *Journal of Business Continuity & Emergency Planning*, *15*(2), 158–170.

27. Masson, A. (2014). The extratropical transition of Hurricane Igor and the impacts on Newfoundland. *Natural Hazards*, *72*(2), 617–632. https://doi.org/10.1007/s11069-013-1027-x

28. Straub, A. M. (2024). Post-tropical cyclone Fiona and Atlantic Canada: Media framing of hazard risk in the Anthropocene. *Disasters*, *48*(4), e12641. https://doi.org/10.1111/disa.12641

29. Taylor, A. R., Dracup, E., MacLean, D. A., Boulanger, Y., & Endicott, S. (2019). Forest structure more important than topography in determining windthrow during Hurricane Juan in Canada's Acadian Forest. *Forest Ecology and Management*, *434*, 255–263. https://doi.org/10.1016/j.foreco.2018.12.026

30. Boluwade, A., & Farooque, A. A. (2024). Spatial and conventional verifications of hurricanes Dorian and Fiona using the Canadian precipitation analysis & integrated multi-satellite retrievals for GPM products. *Journal of Hydrology*, *639*, 131611. https://doi.org/10.1016/j.jhydrol.2024.131611

31. Jardine, D. E., Wang, X., & Fenech, A. L. (2021). Highwater Mark Collection after Post Tropical Storm Dorian and Implications for Prince Edward Island, Canada. *Water*, *13*(22), Article 22. https://doi.org/10.3390/w13223201

32. Mulligan, R. P., Swatridge, L., Cantelon, J. A., Kurylyk, B. L., George, E., & Houser, C. (2023). Local and Remote Storm Surge Contributions to Total Water Levels in the Gulf of St. Lawrence During Hurricane Fiona. *Journal of Geophysical Research: Oceans*, *128*(8), e2023JC019910. https://doi.org/10.1029/2023JC019910

33. Cousineau, J., & Murphy, E. (2022). Numerical Investigation of Climate Change Effects on Storm Surges and Extreme Waves on Canada's Pacific Coast. *Atmosphere*, *13*(2), Article 2. https://doi.org/10.3390/atmos13020311

34. Jiang, J., & Perrie, W. (2008). Climate change effects on North Atlantic cyclones. *Journal of Geophysical Research: Atmospheres*, *113*(D9). https://doi.org/10.1029/2007JD008749

35. Li, S. H., Irwin, P., Lounis, Z., Attar, A., Dale, J., Gibbons, M., & Beaulieu, S. (2022). Effects of Nonstationarity of Extreme Wind Speeds and Ground Snow Loads in a Future Canadian Changing Climate. *Natural Hazards Review*, *23*(4), 04022022. https://doi.org/10.1061/(ASCE)NH.1527-6996.0000572

36. Trepanier, J. C. (2020). North Atlantic Hurricane Winds in Warmer than Normal Seas. *Atmosphere*, *11*(3), Article 3. https://doi.org/10.3390/atmos11030293

37. Rana, A., Zhu, Q., Detken, A., Whalley, K., & Castet, C. (2022). Strengthening climate-resilient development and transformation in Viet Nam. *Climatic Change*, *170*(1), 4. https://doi.org/10.1007/s10584-021-03290-y

38. Ryan, B., & Bristow, D. N. (2024). Risk assessment framework of adapting coastal infrastructure to climate change. *AIP Conference Proceedings*, *3215*(1), 020009. https://doi.org/10.1063/5.0241230

39. Do, C., & Kuleshov, Y. (2023). Multi-Hazard Tropical Cyclone Risk Assessment for Australia. *Remote Sensing*, *15*(3), Article 3. https://doi.org/10.3390/rs15030795

40. Forzieri, G., Feyen, L., Russo, S., Vousdoukas, M., Alfieri, L., Outten, S., Migliavacca, M., Bianchi, A., Rojas, R., & Cid, A. (2016). Multi-hazard assessment in Europe under climate change. *Climatic Change*, *137*(1), 105–119. https://doi.org/10.1007/s10584-016-1661-x

41. Liu, A. (2024). Integrated Multi-Hazard and Vulnerability Modelling for Flood Risk Assessment in the US Gulf Coast [Masters, Department of Risk and Disaster Reduction]. In *Masters thesis, Department of Risk and Disaster Reduction*. https://discovery.ucl.ac.uk/id/eprint/10198115/

42. Sahoo, B., & Bhaskaran, P. K. (2018). Multi-hazard risk assessment of coastal vulnerability from tropical cyclones – A GIS based approach for the Odisha coast. *Journal of Environmental Management*, *206*, 1166–1178. https://doi.org/10.1016/j.jenvman.2017.10.075

43. Chouinard, O., Plante, S., Weissenberger, S., Noblet, M., & Guillemot, J. (2017). The Participative Action Research Approach to Climate Change Adaptation in Atlantic Canadian Coastal Communities. In W. Leal Filho & J. M. Keenan (Eds.), *Climate Change Adaptation in North America: Fostering Resilience and the Regional Capacity to Adapt* (pp. 67–87). Springer International Publishing. https://doi.org/10.1007/978-3-319-53742-9_5

44. Manuel, P., Rapaport, E., Keefe, J., & Krawchenko, T. (2015). Coastal climate change and aging communities in Atlantic Canada: A methodological overview of community asset and social vulnerability mapping. *Canadian Geographies / Géographies Canadiennes*, 59(4), 433–446. https://doi.org/10.1111/cag.12203

45. Provan, M., Ferguson, S., & Murphy, E. (2022). Storm surge contributions to flood hazards on Canada's Atlantic Coast. *Journal of Flood Risk Management*, 15(3), e12800. https://doi.org/10.1111/jfr3.12800

46. Thompson, K. R., Bernier, N. B., & Chan, P. (2009). Extreme sea levels, coastal flooding and climate change with a focus on Atlantic Canada. *Natural Hazards*, 51(1), 139–150. https://doi.org/10.1007/s11069-009-9380-5

47. Vasseur, L., Thornbush, M., & Plante, S. (2017). Climatic and Environmental Changes Affecting Communities in Atlantic Canada. *Sustainability*, 9(8), Article 8. https://doi.org/10.3390/su9081293

48. Carney, M., Kantz, H., & Nicol, M. (2022). *Hurricane Simulation and Nonstationary Extremal Analysis for a Changing Climate*. https://doi.org/10.1175/JAMC-D-22-0003.1

49. Hillier, J. K., Matthews, T., Wilby, R. L., & Murphy, C. (2020). Multi-hazard dependencies can increase or decrease risk. *Nature Climate Change*, 10(7), 595–598. https://doi.org/10.1038/s41558-020-0832-y

50. Nofal, O. M., van de Lindt, J. W., Do, T. Q., Yan, G., Hamideh, S., Cox, D. T., & Dietrich, J. C. (2021). Methodology for Regional Multihazard Hurricane Damage and Risk Assessment. *Journal of Structural Engineering*, 147(11), 04021185. https://doi.org/10.1061/(ASCE)ST.1943-541X.0003144

51. Stalhandske, Z., Steinmann, C. B., Meiler, S., Sauer, I. J., Vogt, T., Bresch, D. N., & Kropf, C. M. (2024). Global multi-hazard risk assessment in a changing climate. *Scientific Reports*, 14(1), 5875. https://doi.org/10.1038/s41598-024-55775-2

52. Tilloy, A., Malamud, B. D., Winter, H., & Joly-Laugel, A. (2019). A review of quantification methodologies for multi-hazard interrelationships. *Earth-Science Reviews*, 196, 102881. https://doi.org/10.1016/j.earscirev.2019.102881

53. Ward, P. J., Daniell, J., Duncan, M., Dunne, A., Hananel, C., Hochrainer-Stigler, S., Tijssen, A., Torresan, S., Ciurean, R., Gill, J. C., Sillmann, J., Couasnon, A., Koks, E., Padrón-Fumero, N., Tatman, S., Tronstad Lund, M., Adesiyun, A., Aerts, J. C. J. H., Alabaster, A., … de Ruiter, M. C. (2022). Invited perspectives: A research agenda towards disaster risk management pathways in multi-(hazard-)risk assessment. *Natural Hazards and Earth System Sciences*, 22(4), 1487–1497. https://doi.org/10.5194/nhess-22-1487-2022

54. Kameshwar, S., & Padgett, J. E. (2014). Multi-hazard risk assessment of highway bridges subjected to earthquake and hurricane hazards. *Engineering Structures*, 78, 154–166. https://doi.org/10.1016/j.engstruct.2014.05.016

55. Román, M. O., Wang, Z., Sun, Q., Kalb, V., Miller, S. D., Molthan, A., Schultz, L., Bell, J., Stokes, E. C., Pandey, B., Seto, K. C., Hall, D., Oda, T., Wolfe, R. E., Lin, G., Golpayegani, N., Devadiga, S., Davidson, C., Sarkar, S., … Masuoka, E. J. (2018). NASA's Black Marble nighttime lights product suite. *Remote Sensing of Environment*, 210, 113–143. https://doi.org/10.1016/j.rse.2018.03.017

56. Eberenz, S., Lüthi, S., & Bresch, D. N. (2021). Regional tropical cyclone impact functions for globally consistent risk assessments. *Natural Hazards and Earth System Sciences*, 21(1), 393–415. https://doi.org/10.5194/nhess-21-393-2021

57. Emanuel, K. (2011). *Global Warming Effects on U.S. Hurricane Damage*. https://doi.org/10.1175/WCAS-D-11-00007.1

58. Emanuel, K. (2022). *Tropical Cyclone Seeds, Transition Probabilities, and Genesis*. https://doi.org/10.1175/JCLI-D-21-0922.1

59. Emanuel, K., Ravela, S., Vivant, E., & Risi, C. (2006). *A Statistical Deterministic Approach to Hurricane Risk Assessment*. https://doi.org/10.1175/BAMS-87-3-299

60. Lin, J., Rousseau-Rizzi, R., Lee, C.-Y., & Sobel, A. (2023). An Open-Source, Physics-Based, Tropical Cyclone Downscaling Model With Intensity-Dependent Steering. *Journal of Advances in Modeling Earth Systems*, 15(11), e2023MS003686. https://doi.org/10.1029/2023MS003686

61. Emanuel, K. (2017). A fast intensity simulator for tropical cyclone risk analysis. *Natural Hazards*, 88(2), 779–796. https://doi.org/10.1007/s11069-017-2890-7

62. Hall, A., Rahimi, S., Norris, J., Ban, N., Siler, N., Leung, L. R., Ullrich, P., Reed, K. A., Prein, A. F., & Qian, Y. (2024). An Evaluation of Dynamical Downscaling Methods Used to Project Regional Climate Change. *Journal of Geophysical Research: Atmospheres*, 129(24), e2023JD040591. https://doi.org/10.1029/2023JD040591

63. Knutson, T. R., Sirutis, J. J., Garner, S. T., Vecchi, G. A., & Held, I. M. (2008). Simulated reduction in Atlantic hurricane frequency under twenty-first-century warming conditions. *Nature Geoscience*, *1*(6), 359–364. https://doi.org/10.1038/ngeo202

64. Prein, A. F., Liu, C., Ikeda, K., Trier, S. B., Rasmussen, R. M., Holland, G. J., & Clark, M. P. (2017). Increased rainfall volume from future convective storms in the US. *Nature Climate Change*, *7*(12), 880–884. https://doi.org/10.1038/s41558-017-0007-7

65. Rasmussen, R., Liu, C., Ikeda, K., Gochis, D., Yates, D., Chen, F., Tewari, M., Barlage, M., Dudhia, J., Yu, W., Miller, K., Arsenault, K., Grubišić, V., Thompson, G., & Gutmann, E. (2011). *High-Resolution Coupled Climate Runoff Simulations of Seasonal Snowfall over Colorado: A Process Study of Current and Warmer Climate*. https://doi.org/10.1175/2010JCLI3985.1

66. Schär, C., Frei, C., Lüthi, D., & Davies, H. C. (1996). Surrogate climate-change scenarios for regional climate models. *Geophysical Research Letters*, *23*(6), 669–672. https://doi.org/10.1029/96GL00265

67. Willison, J., Robinson, W. A., & Lackmann, G. M. (2015). *North Atlantic Storm-Track Sensitivity to Warming Increases with Model Resolution*. https://doi.org/10.1175/JCLI-D-14-00715.1

68. Holland, G. (1980). An Analytic Model of the Wind and Pressure Profiles in Hurricanes. *Mon. Weather Rev.*, *108*, 1212–1218. https://doi.org/10.1175/1520-0493(1980)108<1212:AAMOTW>2.0.CO;2

69. Snaiki, R., & Wu, T. (2017). Modeling tropical cyclone boundary layer: Height-resolving pressure and wind fields. *Journal of Wind Engineering and Industrial Aerodynamics*, *170*, 18–27. https://doi.org/10.1016/j.jweia.2017.08.005

70. Meng, Y., Matsui, M., & Hibi, K. (1995). An analytical model for simulation of the wind field in a typhoon boundary layer. *Journal of Wind Engineering and Industrial Aerodynamics*, *56*(2), 291–310. https://doi.org/10.1016/0167-6105(94)00014-5

71. Lin, N., & Chavas, D. (2012). On hurricane parametric wind and applications in storm surge modeling. *Journal of Geophysical Research: Atmospheres*, *117*(D9). https://doi.org/10.1029/2011JD017126

72. Harper, B., Kepert, J., & Ginger, J. (2010). *Guidelines for converting between various wind averaging periods in tropical cyclone conditions*.

73. Didier, D., Baudry, J., Bernatchez, P., Dumont, D., Sadegh, M., Bismuth, E., Bandet, M., Dugas, S., & Sévigny, C. (2019). Multihazard simulation for coastal flood mapping: Bathtub versus numerical modelling in an open estuary, Eastern Canada. *Journal of Flood Risk Management*, *12*(S1), e12505. https://doi.org/10.1111/jfr3.12505

74. Poulter, B., & Halpin, P. N. (2008). Raster modelling of coastal flooding from sea-level rise. *International Journal of Geographical Information Science*, *22*(2), 167–182. https://doi.org/10.1080/13658810701371858

75. Shepard, C. C., Agostini, V. N., Gilmer, B., Allen, T., Stone, J., Brooks, W., & Beck, M. W. (2012). Assessing future risk: Quantifying the effects of sea level rise on storm surge risk for the southern shores of Long Island, New York. *Natural Hazards*, *60*(2), 727–745. https://doi.org/10.1007/s11069-011-0046-8

76. Xu, L. (2010). A simple coastline storm surge model based on pre-run SLOSH outputs. *29th Conference on Hurricanes and Tropical Meteorology (10-14 May 2010)*.

77. Makkonen, L. (2006). *Plotting Positions in Extreme Value Analysis*. https://doi.org/10.1175/JAM2349.1

78. Weibull, W. (1939). A statistical theory of strength of materials. *IVB-Handl.*

79. Aznar-Siguan, G., & Bresch, D. N. (2019). CLIMADA v1: A global weather and climate risk assessment platform. *Geoscientific Model Development*, *12*(7), 3085–3097. https://doi.org/10.5194/gmd-12-3085-2019

80. Hart, R. E., & Evans, J. L. (2001). *A Climatology of the Extratropical Transition of Atlantic Tropical Cyclones*. https://journals.ametsoc.org/view/journals/clim/14/4/1520-0442_2001_014_0546_acotet_2.0.co_2.xml

81. Buttle, J. M., Allen, D. M., Caissie, D., Davison, B., Hayashi, M., Peters, D. L., Pomeroy, J. W., Simonovic, S., St-Hilaire, A., & Whitfield, P. H. (2016). Flood processes in Canada: Regional and special aspects. *Canadian Water Resources Journal / Revue Canadienne Des Ressources Hydriques*, *41*(1–2), 7–30. https://doi.org/10.1080/07011784.2015.1131629

82. Lin, H., Mo, R., Vitart, F., & Stan, C. (2019). Eastern Canada Flooding 2017 and its Subseasonal Predictions. *Atmosphere-Ocean*, *57*(3), 195–207. https://doi.org/10.1080/07055900.2018.1547679

83. Hersbach, H., Bell, B., Berrisford, P., Hirahara, S., Horányi, A., Muñoz-Sabater, J., Nicolas, J., Peubey, C., Radu, R., Schepers, D., Simmons, A., Soci, C., Abdalla, S., Abellan, X., Balsamo, G., Bechtold, P., Biavati, G., Bidlot, J., Bonavita, M., … Thépaut, J.-N. (2020). The ERA5 global reanalysis. *Quarterly Journal of the Royal Meteorological Society*, *146*(730), 1999–2049. https://doi.org/10.1002/qj.3803

84. Snaiki, R., & Wu, T. (2025). A Novel Dynamic Bias-Correction Framework for Hurricane Risk Assessment under Climate Change. *arXiv Preprint arXiv:2505.00832*.

85. Saviz Naeini, S., Snaiki, R., & Wu, T. (2025). Advancing spatio-temporal storm surge prediction with hierarchical deep neural networks. *Natural Hazards*. https://doi.org/10.1007/s11069-025-07428-4